\documentclass[iop]{emulateapj}
\usepackage[modulo]{lineno}
\usepackage[xcolor]{xcolor}
\usepackage{amsmath}
\usepackage[breaklinks,colorlinks,citecolor=blue]{hyperref} 

\def\deg{{$^{\circ}$}}
\def\bd{\object[BD+44 493]{BD$+$44\deg493}}

\def\kmsec{\mbox{km~s$^{\rm -1}$}}
\def\logg{\mbox{log~{\it g}}}
\def\msun{\mbox{$M_{\odot}$}}
\def\teff{\mbox{$T_{\rm eff}$}}

\def\cempno{\mbox{CEMP-no}}
\def\rpro{\mbox{$r$-process}}

\def\loggf{\mbox{$\log gf$}}

\shorttitle{P, S, and Zn in BD$+$44\deg493}
\shortauthors{Roederer, Placco, \& Beers}
\slugcomment{Accepted for publication in the Astrophysical Journal Letters}

\begin{document}

\title{
Detection of Phosphorus, Sulphur, and Zinc in the \\
Carbon-Enhanced Metal-Poor Star
BD$+$44\deg493\footnotemark[$\dagger$]
}

\footnotetext[$\dagger$]{Based 
on observations made with the NASA/ESA 
\textit{Hubble Space Telescope}, 
obtained at the Space Telescope Science Institute (STScI), which is 
operated by the Association of Universities for 
Research in Astronomy, Inc.\ (AURA) under NASA contract NAS~5-26555.
These observations are associated with program GO-14231.}

\author{
Ian U.\ Roederer,\altaffilmark{1,2}
Vinicius M.\ Placco,\altaffilmark{2,3}
and
Timothy C.\ Beers\altaffilmark{2,3}
}

\altaffiltext{1}{Department of Astronomy, University of Michigan,
1085 S.\ University Ave., Ann Arbor, MI 48109, USA;
\mbox{iur@umich.edu}
}
\altaffiltext{2}{Joint Institute for Nuclear Astrophysics and Center for the
Evolution of the Elements (JINA-CEE), USA
}
\altaffiltext{3}{Department of Physics, University of Notre Dame,
Notre Dame, IN 46556, USA
}


\addtocounter{footnote}{3}

\begin{abstract}

The carbon-enhanced metal-poor star
BD$+$44\deg493
([Fe/H]~$= -$3.9)
has been proposed as a candidate second-generation star
enriched by metals from a single Pop~III star.
We report the first detections of P and S
and the second detection of Zn
in any extremely metal-poor carbon-enhanced star,
using new spectra of BD$+$44\deg493
collected by the Cosmic Origins Spectrograph
on the \textit{Hubble Space Telescope}.
We derive
[P/Fe]~$= -$0.34~$\pm$~0.21,
[S/Fe]~$= +$0.07~$\pm$~0.41, and
[Zn/Fe]~$= -$0.10~$\pm$~0.24.
We increase by ten-fold the number of
Si~\textsc{i} lines detected in 
BD$+$44\deg493,
yielding 
[Si/Fe]~$= +$0.15~$\pm$~0.22.
The [S/Fe] and [Zn/Fe] ratios 
exclude the hypothesis that
the abundance pattern in 
BD$+$44\deg493 results from depletion
of refractory elements 
onto dust grains.
Comparison with 
zero-metallicity supernova models
suggests that the 
stellar progenitor that enriched BD$+$44\deg493
was massive and ejected much less than 
0.07~\msun\ of $^{56}$Ni,
characteristic of a faint supernova.


\end{abstract}


\section{Introduction}
\label{intro}

Carbon-enhanced metal-poor stars with no 
enhancements of $n$-capture elements
(\cempno; \citealt{beers05}) 
may have formed from the yields of
only a single prior generation of zero-metallicity 
Pop~III stars, 
and perhaps only one star. 
\citet{norris13} summarize this evidence for this hypothesis, 
linking carbon enhancement and
low metallicity with remote,
chemically-primitive regions 
that experienced few enrichment events 
(e.g., \citealt{frebel10,cooke11,carollo12}).
Comparing model yields to the
detailed abundance patterns observed in \cempno\
stars provides a rare opportunity
to constrain the nature and mechanics
of low- or zero-metallicity supernovae (SNe).
Stars that produce high C/Fe ratios in the early Universe
may include massive, rapidly-rotating, zero-metallicity
stars
\citep{meynet06} and
SNe or hypernovae
that experienced 
mixing and fallback explosions
(e.g., \citealt{umeda03,ishigaki14}).

There are approximately 70 confirmed members 
of the class of \cempno\ stars \citep{placco14b},
but only one, \bd,
is sufficiently bright ($V =$~9.1)
to readily observe in the ultraviolet (UV)
with the echelle spectrographs on the
\textit{Hubble Space Telescope} (\textit{HST}).~
\bd\ is a red giant
located $\sim$~200~pc from the Sun
on a typical halo orbit \citep{ito13}.
It shows no evidence of stellar companions,
as radial velocity measurements
constant to within $\approx$~0.6~\kmsec\
span more than 31~years
(\citealt{carney03}; \citealt{hansen16}, and references therein).
\bd\ is 
extremely metal-poor ([Fe/H]~$= -$3.88~$\pm$~0.19),
carbon-enhanced ([C/Fe]~$= +$1.23~$\pm$~0.20),
and exhibits low levels of
elements
produced by $n$-capture reactions
(e.g., [Ba/Fe]~$= -$0.60~$\pm$~0.12).
This fortuitous combination of characteristics
was first identified by \citet{ito09}.
\citet{ito13} performed a detailed
abundance analysis of \bd\ using
high-quality optical spectra. 
\citet{takeda13}, \citet{placco14a}, 
\citet{roederer14a},
and
\citet{aoki15} 
confirmed the C-enhanced nature of \bd.
\citet{placco14a}\ presented the first
abundance analysis of the UV
spectrum (2280--3100~\AA) of \bd,
which included several elements
not detectable in optical spectra.

Here, we report the detection of
phosphorus (P, $Z =$~15),
sulphur (S, $Z =$~16), and 
zinc (Zn, $Z =$~30) in \bd.
These are the first detections of P and S
and the second detection of Zn
in any \cempno\ star with [Fe/H]~$\lesssim -$4.
Expanding the inventory of elements 
for even a single member of the class of \cempno\
stars provides critical insight 
into at least one of the likely progenitors 
of such stars in the early Universe.

\section{Observations}
\label{observations}

We have obtained new observations of portions of the
UV spectrum of \bd\ using 
the Cosmic Origins Spectrograph (COS; \citealt{green12})
on \textit{HST}.~
Three medium-resolution grating 
settings were used to collect 
spectra covering $\sim$~270~\AA\ 
between 1799 and 2348~\AA\ in
nine non-contiguous stripes of $\sim$~30~\AA\ each.
The spectral resolving power,
measured from the PtNe comparison lamp spectra,
ranges over 13,000~$< R <$~17,000.
Table~\ref{obstab} presents a log of these observations,
including the date,
exposure time ($t_{\rm exp}$),
grating setting,
central wavelength setting ($\lambda_{\rm c}$),
wavelengths observed ($\lambda$), and 
signal-to-noise ratios (S/N) per pixel
after all observations have been co-added.

\begin{deluxetable}{lccccc}
\tablecaption{Log of Observations
\label{obstab}}
\tablewidth{0pt}
\tabletypesize{\scriptsize}
\tablehead{
\colhead{Date} &
\colhead{$t_{\rm exp}$} &
\colhead{Grating} &
\colhead{$\lambda_{\rm c}$} &
\colhead{$\lambda$\tablenotemark{a}} &
\colhead{S/N} \\
\colhead{} &
\colhead{(s)} &
\colhead{} &
\colhead{(\AA)} &
\colhead{(\AA)} &
\colhead{} 
}
\startdata
2015 Nov 21       & 8417 & G225M & 2233 & 2120--2153 & 20 \\
                  &      &       &      & 2222--2249 & 40 \\
                  &      &       &      & 2320--2348 & 55 \\
2015 Nov 22       & 8301 & G185M & 1971 & 1855--1886 & 12 \\
                  &      &       &      & 1957--1987 & 20 \\
                  &      &       &      & 2060--2090 & 30 \\
2015 Dec 04,08,09 &31035 & G185M & 1913 & 1799--1829 & 10 \\
                  &      &       &      & 1902--1932 & 35 \\
                  &      &       &      & 2006--2034 & 55 
\enddata
\tablecomments{All observations are
associated with datasets
LCTQ01010--06010 
in the Mikulski Archive for Space Telescopes (MAST).}
\tablenotetext{a}{Air wavelengths are given for
$\lambda >$~2000~\AA\ and vacuum values below.
Values given are corrected to rest wavelengths.}
\end{deluxetable}

\section{Analysis}
\label{analysis}

We adopt the model atmosphere parameters for \bd\
derived by \citet{ito13}:\
\teff, 5430~$\pm$~150~K;
\logg, 3.4~$\pm$~0.3;
microturbulence velocity, 1.3~$\pm$~0.3~\kmsec;
and
[M/H], $-$3.8~$\pm$~0.2.
We use a 1D plane-parallel
model interpolated from the ATLAS9
$\alpha$-enhanced grid of \citet{castelli03}, and
we derive abundances using a recent version of the
spectrum analysis code MOOG \citep{sneden73,sobeck11}.

\begin{figure}
\begin{center}
\includegraphics[angle=0,width=3.4in]{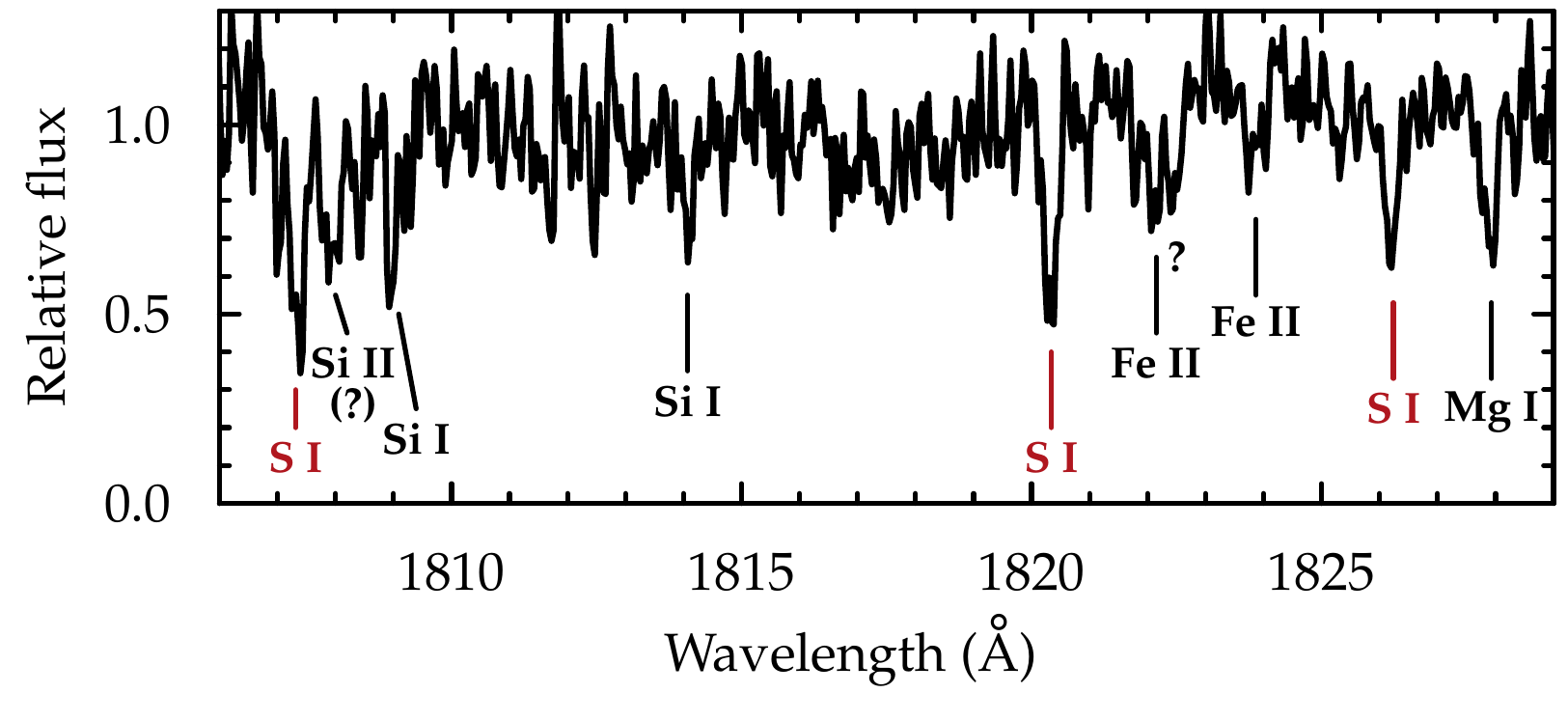} \\
\vspace*{-0.0in}
\includegraphics[angle=0,width=3.4in]{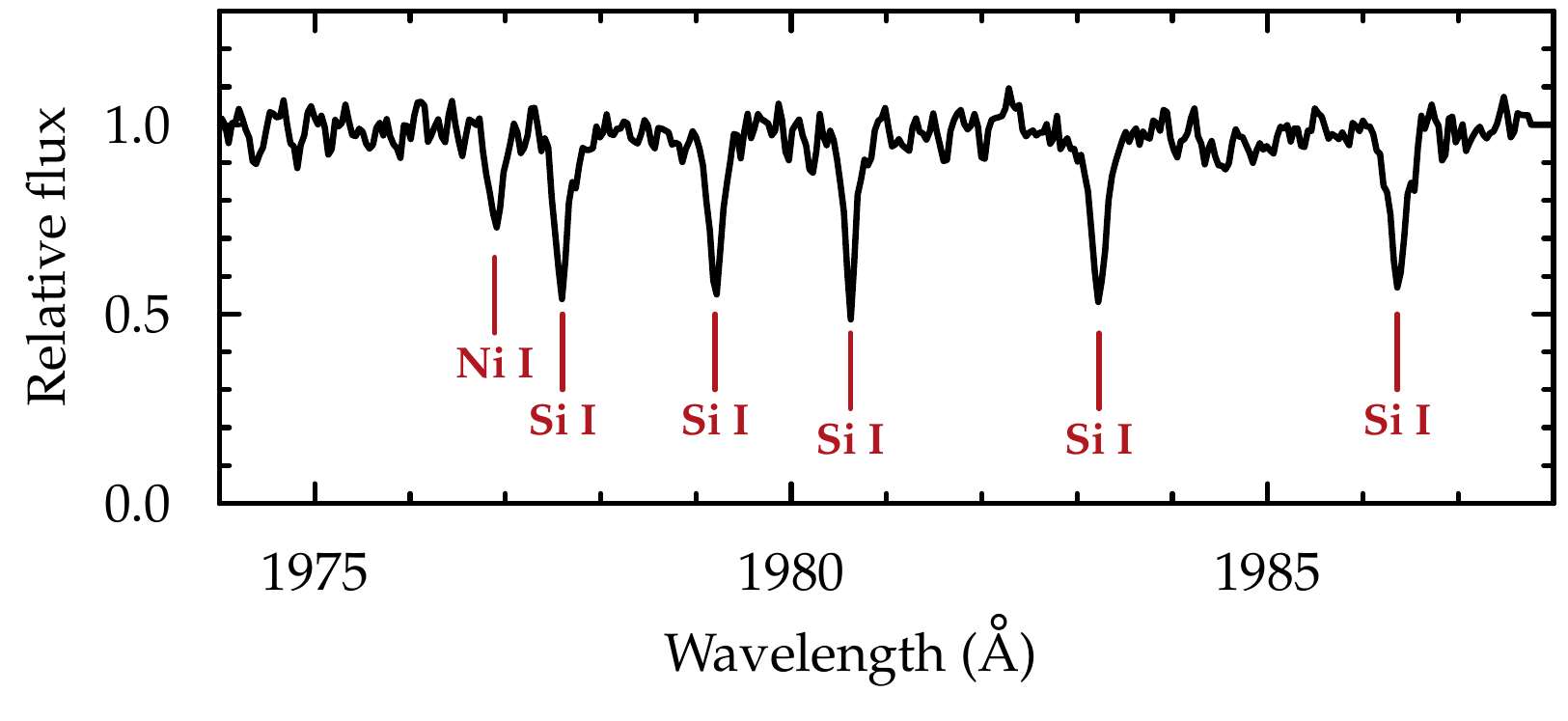} \\
\vspace*{-0.0in}
\includegraphics[angle=0,width=3.4in]{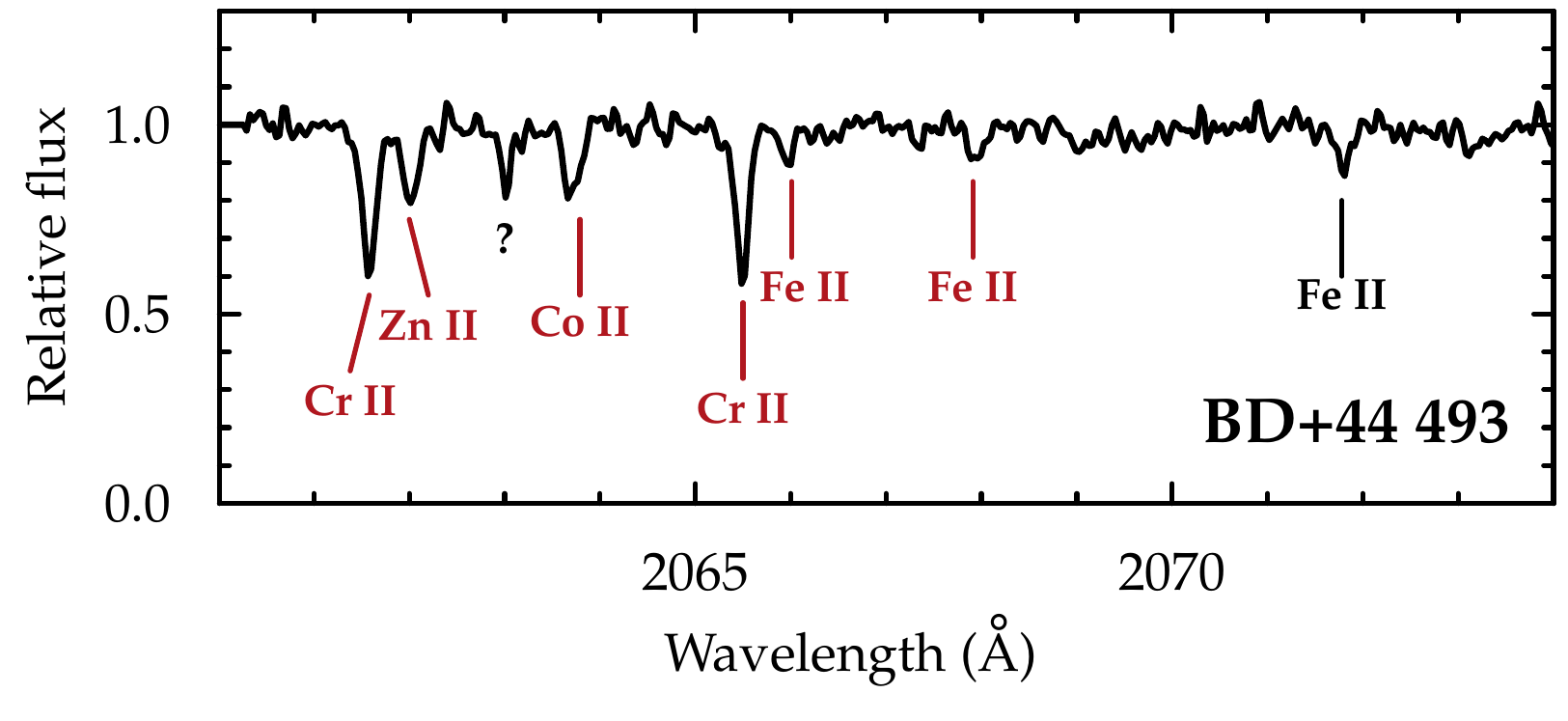} \\
\vspace*{-0.0in}
\includegraphics[angle=0,width=3.4in]{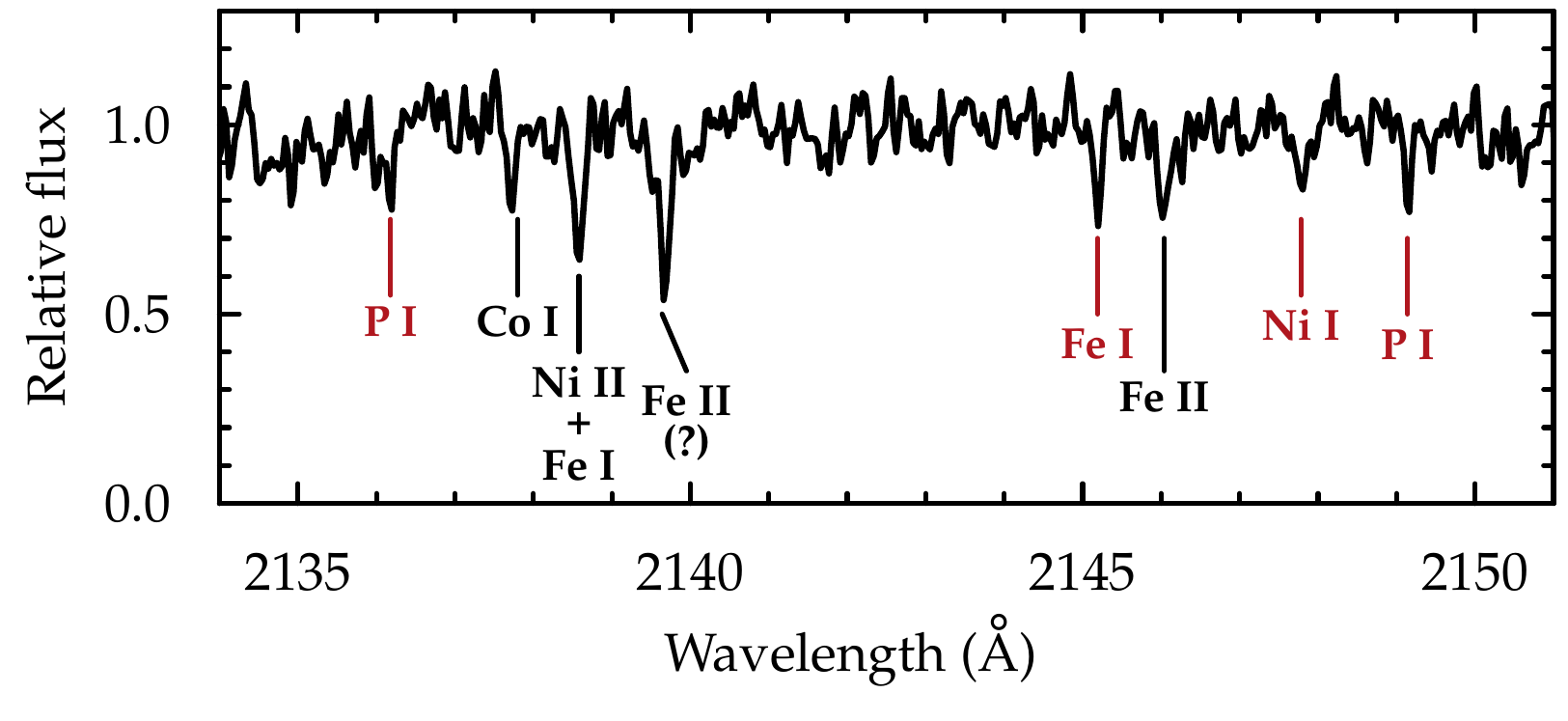} 
\end{center}
\caption{
\label{specplot}
Sections of the COS spectra of \mbox{BD$+$44\deg493}.
Lines 
that have been used to derive abundances
are marked in red.
Uncertain or unknown line identifications
are labeled with a question mark;
these lines are not used and do not
adversely affect our analysis.}
\end{figure}

Figure~\ref{specplot}
shows sections of the COS spectra of \bd.
Note that 
the continuum is visible
and the lines of interest are
unblended.
We measure equivalent widths (EW)
by fitting the observed lines 
with a convolution of a Gaussian and
the COS line spread function
(LSF; \citealt{ghavamian09}).
Part of our COS spectrum overlaps (2320--2348~\AA)
with the Space Telescope Imaging Spectrograph
(STIS) spectrum analyzed by \citet{placco14a},
and we compare EWs in this region to check
our measurements.
Failure to account for the COS LSF,
which redistributes absorption from the line core
to the wings, leads to an underestimate 
of the EWs by $\sim$~25\%.
When the COS LSF is included, 
the EWs measured from the COS and STIS spectra
agree to better than 5\%.
We regard 
the standard deviation of the residuals, 6.8~m\AA,
as the minimum uncertainty in 
our EWs.

\section{Results}
\label{results}

The 
EWs, atomic data, and 
abundances derived for each line are listed in
Table~\ref{abundtab}.
Table~\ref{meantab}
lists the mean abundances.
We calculate uncertainties
as described in \citet{roederer14b}.
We adopt the
Solar reference abundances
given in \citet{asplund09}.

\begin{deluxetable}{ccccccc}
\tablecaption{Atomic Data, Equivalent Widths, and 
Abundances Derived from Individual Lines
\label{abundtab}}
\tablewidth{0pt}
\tabletypesize{\scriptsize}
\tablehead{
\colhead{Spec.} &
\colhead{$\lambda$\tablenotemark{a}} &
\colhead{E.P.} &
\colhead{\loggf} &
\colhead{Ref.} &
\colhead{EW} &
\colhead{$\log\epsilon$} \\
\colhead{} &
\colhead{(\AA)} &
\colhead{(eV)} &
\colhead{} &
\colhead{} &
\colhead{(m\AA)} &
\colhead{}
}
\startdata
Mg~\textsc{i}  & 2025.82 &  0.00 & $-$0.95 (0.03) & 1 & 208.4 & 4.23 (0.50)  \\ 
Si~\textsc{i}  & 1874.84 &  0.78 & $-$0.77 (0.30) & 1 & 107.7 & 4.05 (0.46)  
\enddata
\tablerefs{1:\ NIST; 
2:\ \citet{bergeson93};
3:\ \citet{wood14};
4:\ \citet{fedchak99};
5:\ \citet{roederer12a}}
\tablenotetext{a}{Air wavelengths are given for 
$\lambda >$~2000~\AA\ and vacuum values below.}
\tablecomments{The complete version of Table~\ref{abundtab} is
available in the online edition of the Journal. 
A short version is shown here to illustrate its form and content.}
\end{deluxetable}

\begin{deluxetable}{cccccc}
\tablecaption{Mean Abundances and Comparison with Previous Results
\label{meantab}}
\tablewidth{0pt}
\tabletypesize{\scriptsize}
\tablehead{
\colhead{Spec.} &
\colhead{N} &
\colhead{$\log\epsilon$} &
\colhead{[X/Fe]} &
\colhead{[X/Fe]} &
\colhead{[X/Fe]} \\
\colhead{} &
\colhead{} &
\colhead{(Ref.\ 1)} &
\colhead{(Ref.\ 1)} &
\colhead{(Ref.\ 2)} &
\colhead{(Ref.\ 3)} 
}
\startdata
Mg~\textsc{i}  &  1 & 4.23\,(0.50) & $+$0.51\,(0.45) & \nodata         & $+$0.51\,(0.08) \\
Si~\textsc{i}  & 10 & 3.78\,(0.24) & $+$0.15\,(0.22) & \nodata         & $+$0.54\,(0.14) \\
P~\textsc{i}   &  3 & 1.19\,(0.28) & $-$0.34\,(0.21) & \nodata         & \nodata         \\
S~\textsc{i}   &  3 & 3.31\,(0.45) & $+$0.07\,(0.41) & \nodata         & $<+$1.11        \\
Ti~\textsc{ii} &  2 & 1.37\,(0.28) & $+$0.30\,(0.22) & $+$0.36\,(0.20) & $+$0.41\,(0.08) \\
Cr~\textsc{ii} &  2 & 1.99\,(0.23) & $+$0.23\,(0.21) & $-$0.09\,(0.21) & $-$0.17\,(0.10) \\
Fe~\textsc{i}  &  3 & 3.43\,(0.28) & $-$4.07\,(0.28) & $-$3.88\,(0.19) & $-$3.83\,(0.19) \\
Fe~\textsc{ii} & 12 & 3.78\,(0.24) & $-$3.72\,(0.24) & $-$3.87\,(0.17) & $-$3.82\,(0.15) \\
Co~\textsc{ii} &  4 & 1.76\,(0.29) & $+$0.65\,(0.23) & \nodata         & $+$0.59\,(0.06)\tablenotemark{a} \\
Ni~\textsc{i}  &  3 & 2.26\,(0.26) & $-$0.08\,(0.21) & $-$0.02\,(0.22) & $+$0.13\,(0.06) \\
Ni~\textsc{ii} &  3 & 2.27\,(0.23) & $-$0.07\,(0.21) & $-$0.13\,(0.23) & \nodata         \\
Zn~\textsc{ii} &  1 & 0.58\,(0.32) & $-$0.10\,(0.24) & \nodata         & $<+$0.22        \\
As~\textsc{i}  &  1 & $< -$0.21    & $< +$1.37       & \nodata         & \nodata         \\
Se~\textsc{i}  &  3 & $< -$0.07    & $< +$0.47       & \nodata         & \nodata        
\enddata
\tablecomments{All [X/Fe] ratios are computed using [Fe/H]~$= -$3.88.
[Fe/H] is listed for Fe~\textsc{i} and Fe~\textsc{ii}.}
\tablerefs{1:\ This study;
2:\ \citet{placco14a};
3:\ \citet{ito13}}
\tablenotetext{a}{[Co~\textsc{i}/Fe]}
\end{deluxetable}

Table~\ref{meantab} compares our abundances
with those derived by previous studies.
Our Fe abundances 
are in agreement with those derived
from other optical and UV lines,
indicating that there are no
significant zero-point differences.
The [Mg/Fe], [Ti/Fe], [Co/Fe], and [Ni/Fe]
ratios are also in agreement with previous work.
The [Cr/Fe] we derive from Cr~\textsc{ii} lines
agrees
with the \citet{placco14a} value,
but it is higher than the \citet{ito13} value.
If we adopt
the \loggf\ scale from \citet{nilsson06},
whose \loggf\ values for the two Cr~\textsc{ii} lines
examined by us agree to within 0.02~dex
with the \citet{bergeson93} \loggf\ values,
the \citeauthor{ito13}\ [Cr~\textsc{ii}/Fe] 
ratio would increase by $\approx$~0.2~dex.
This resolves the discrepancy.

We have detected several Co~\textsc{ii} lines,
which may be desaturated by
hyperfine splitting (HFS;
\citealt{lawler15}).
Laboratory data to reconstruct the HFS are
not available at present, 
but the concordance between the [Co/Fe] ratios
derived from Co~\textsc{i} and Co~\textsc{ii} lines
(Table~\ref{meantab})
suggests we are not overestimating the
Co abundance by neglecting Co~\textsc{ii} HFS
(cf.\ \citealt{sneden16}).

Deviations from local thermodynamic equilibrium (i.e., non-LTE)
are a persistent concern in abundance work.
The Ti~\textsc{ii}, Cr~\textsc{ii}, and Zn~\textsc{ii} 
lines we use
meet the ``gold standard'' criteria \citep{lawler13}:\
unsaturated lines arising from the
ground and low-lying levels of the ion.
These levels are the main population reservoirs
of Fe-group atoms, and they should be well-characterized by LTE.~
The P~\textsc{i} and S~\textsc{i} lines
we use 
arise from low-lying levels that comprise
$\sim$~3--55\% of the neutral species,
which are the dominant ionization states,
so non-LTE overionization
is expected to be minimal.

\section{New Abundances
in BD$+$44\deg493}
\label{newabunds}

We have derived abundances from 10~Si~\textsc{i} lines,
whereas previously only a single optical Si~\textsc{i} line
had been detected in \bd.
We derive [Si/Fe]~$= +$0.15~$\pm$~0.22.
The statistical error in the Si abundance is only 0.06~dex;
the error in [Si/Fe] is dominated
by uncertainties in Fe and the model atmosphere parameters.
Our derived [Si/Fe] ratio 
is lower
than that derived previously
($+$0.54~$\pm$~0.14; \citealt{ito13}).

Our P abundance is 0.6~dex lower
than the upper limit derived by 
\citet{roederer14c}.
The sub-solar [P/Fe] ratio in \bd\ ($-$0.34~$\pm$~0.21)
is also
lower than the mean [P/Fe] ratio
($+$0.04~$\pm$~0.10)
found by \citeauthor{roederer14c}\ for seven 
metal-poor carbon-normal stars.
This could suggest a different origin for the P in \bd,
although a larger comparison sample is highly desirable.

Our S abundance is 0.9~dex 
lower than the upper limits derived by
\citet{takeda11} and \citet{ito13}.
[S/Fe] exhibits the common $\alpha$-enhanced plateau 
in other metal-poor stars
([S/Fe]~$\approx +$0.3; \citealt{nissen07}),
and the [S/Fe] ratio in \bd\
($+$0.07~$\pm$~0.41)
agrees with this value.

The [Zn/Fe] ratio ($-$0.10~$\pm$~0.24)
in \bd\ contrasts with other
stars with [Fe/H]~$\lesssim -$3.3, where
$+$0.4~$\lesssim$~[Zn/Fe]~$\lesssim +$0.7
(e.g., \citealt{nissen07}).
Zn has been detected in one
other \cempno\ star with [Fe/H]~$\sim -$4,
\object[BPS CS 22949-037]{\mbox{CS~22949--037}} \citep{depagne02},
where [Zn/Fe] is enhanced ($+$0.7~$\pm$~0.1).
This suggests diverse conditions for Zn production
in the progenitors of
the \cempno\ stars.

We have also derived 
the first upper limits on the [As/Fe] and [Se/Fe] ratios
for any CEMP star.
These limits indicate that \bd\ is not enhanced 
in elements at the first \rpro\ peak,
similar to carbon-normal 
metal-poor stars \citep{roederer12b}.

\section{Discussion}
\label{Discussion}

The volatile elements C, N, O, S, and Zn 
are among the last 
to condense 
onto dust grains as a molecular cloud cools.
\citet{venn08} proposed depletion 
of refractory elements
on dust in a debris disk 
as an alternative explanation 
for the C enhancement 
and Fe-group depletions 
found in \cempno\ stars. 
While unlikely \citep{venn14},
this hypothesis predicts that
[S/Fe] and [Zn/Fe] in \bd\
should be 
enhanced, similar to [C/Fe], [N/Fe], and [O/Fe].
The solar [S/Fe] and [Zn/Fe] ratios we have derived
exclude this possibility.
The [S/Fe] and [Zn/Fe] ratios 
could also result from a combination of dust depletion and low 
initial ratios produced by the progenitor SN \citep{bonifacio12},
but \bd\ presents evidence against this hypothesis as well.
The mean condensation temperatures ($T_{C}$)
of S and Zn are $\sim$~700~K \citep{lodders03},
roughly half that of Si, Fe, and Ni
($T_{C} \sim$~1350~K).
Si and S should share similar nucleosynthesis histories,
as should Fe, Ni, and Zn.
However, [X/H]~$\sim -$3.9 for all five of these elements
with different $T_{C}$,
arguing against the dust-depletion hypothesis.

We assess the properties of the 
progenitor of \bd\ 
by comparing its abundance pattern with model
predictions for Pop~III SN yields.
We use the 1D,
non-rotating, 
single, massive star models from \citet{heger10}.
These models have masses 
10--100~\msun,
explosion energies
(0.3--10)$\times 10^{51}$~erg,
and vary the amount of mixing in the ejecta.
We simulate 
10$^{4}$ abundance patterns
by resampling the observed ratios from
normal distributions.
We fit all
abundances from C to Zn
using a procedure similar to 
that described in \citet{placco15},
using the \texttt{starfit} 
code\footnote{\href{http://starfit.org}{http://starfit.org}} 
to find the best model
for each resampled pattern.

\begin{figure}
\begin{center}
\includegraphics[angle=0,width=3.4in]{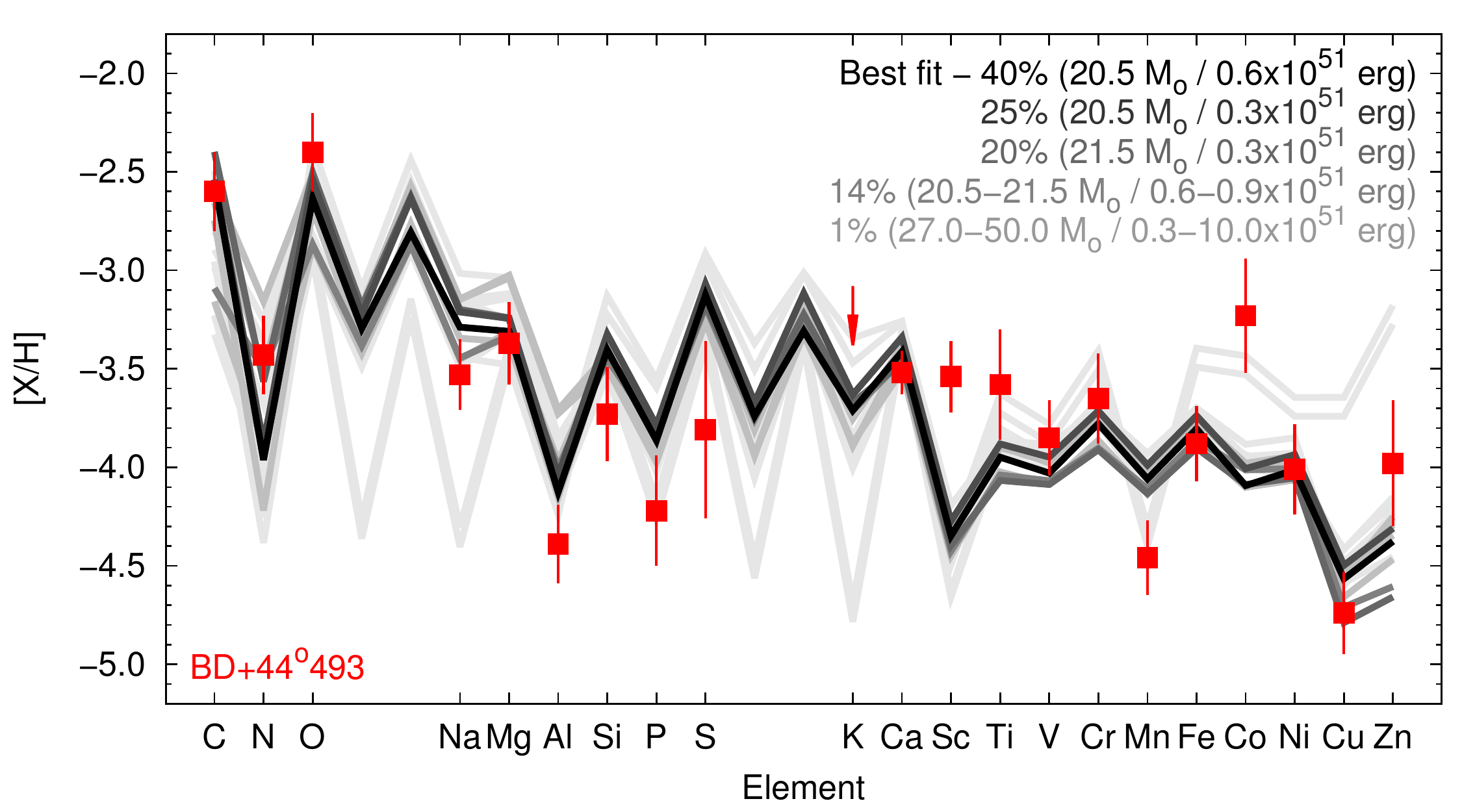}
\end{center}
\caption{
\label{starfit}
Best-fit models for the observed abundances of 
\mbox{BD$+$44\deg493}.
The observational data are shown by red points.
The masses and explosion energies of the models
for the simulated abundance patterns 
are given in the upper right part of each panel,
color-coded by the percentage of their occurrence.
}
\end{figure}

Figure~\ref{starfit} illustrates our results.
The best-fit model,
favored by 40\% of the resampled abundance patterns,
is a progenitor with 
initial mass 20.5~\msun,
explosion energy $0.6\times10^{51}$~erg,
mixing factor $f_{\rm mix} =$~0.1, 
ejected $^{56}$Ni mass 0.0014~\msun, and 
remnant mass 5.47~\msun.
More than 99\% of the favored models
have progenitor masses from 20.5--21.5~\msun\
and explosion energies from (0.3--0.9)$\times 10^{51}$~erg.
Artificially inflating the uncertainties on 
[C/H], [N/H], and [O/H] to 0.4~dex
broadens the range of masses and explosion energies,
but $\sim$~95\%
of the beset-fit
models encompass the same range as the standard case.
\bd\ acquired about 0.001\% of the metals
ejected by this SN, and
the low 
($\ll$~0.07~\msun; \citealt{nomoto06})
ejected $^{56}$Ni mass indicates that
this is a faint SN.
 
The initial mass and explosion energy
are somewhat lower than those favored previously
(25~\msun, 5$\times 10^{51}$~erg; \citealt{ito13}).
This might be expected based on the lower [Zn/Fe] ratio,
which is sensitive to the explosion energy
(e.g., \citealt{umeda05,tominaga07}).
These values are quite model dependent,
and the range of acceptable values
does not represent
the true uncertainties.
The
enhanced [Co/Fe] and solar [Zn/Fe]
cannot be fit simultaneously,
but we note that model fits to \cempno\ stars
often under-predict the [Co/Fe] ratio
(e.g., \citealt{tominaga14}).
A higher progenitor mass and explosion energy
with enhanced mixing
may better explain the enhanced [Co/Zn] ratio
(cf.\ \citealt{nomoto06}),
although this could 
exacerbate discrepancies with other elements
(see discussion in \citealt{tominaga14}).
Massive, rapidly-rotating, low-metallicity stars
have also 
been proposed as the progenitors of \cempno\ stars like \bd\
(e.g., \citealt{maeder15,frischknecht16}),
although no published grids of yields are yet available
for comparison.

Our new observations of P, S, and Zn in \bd\
demonstrate that COS can be used
for abundance work in late-type stars.
This may enable
new theoretical comparisons and 
motivate
observational campaigns to study these
elements in the earliest generations of stars.

\acknowledgments

Generous support 
for Program GO-14231 has been provided by
a grant from STScI,
which is
operated by AURA,
under NASA contract NAS5-26555.
Partial support has also been provided by
grant PHY~14-30152 (Physics Frontier Center/JINA-CEE)
awarded by the U.S.\ National Science Foundation (NSF).~
We thank J.\ Lawler and N.\ Tominaga
for helpful discussions.
We also thank the referee for many helpful suggestions.
This research has made use of NASA's
Astrophysics Data System Bibliographic Services;
the arXiv pre-print server operated by Cornell University;
the SIMBAD and VizieR
database hosted by the
Strasbourg Astronomical Data Center;
the Atomic Spectra Database hosted by 
the National Institute of Standards and Technology;
the MAST at STScI;
IRAF software packages
distributed by the National Optical Astronomy Observatories,
which are operated by AURA,
under cooperative agreement with the NSF;
and the R software package \citep{r}.

{\it Facilities:} 
\facility{HST (COS)}

\end{document}